# Leveraging NFTs for Spectrum Securitization in 6G Networks

Zhixian Zhou
*College of Electronics and Information Engineering*
*Shenzhen University*
Shenzhen, China
zzxsees@163.com

Bin Chen*
*College of Electronics and Information Engineering*
*Shenzhen University*
Shenzhen, China
bchen@szu.edu.cn

Chen Sun
*Wireless Network Research Department*
*Research and Development Center, Sony (China) Limited*
Beijing, China
chen.sun@sony.com

Peichang Zhang
*College of Electronics and Information Engineering*
*Shenzhen University*
Shenzhen, China
pzhang@szu.edu.cn

Shuo Wang
*Wireless Network Research Department*
*Research and Development Center, Sony (China) Limited*
Beijing, China
shuo.wang@sony.com

*Abstract*—Dynamic Spectrum Sharing can enhance spectrum resource utilization by promoting the dynamic distribution of spectrum resources. However, to effectively implement dynamic spectrum resource allocation, certain mechanisms are needed to incentivize primary users to proactively share their spectrum resources. This paper, based on the ERC404 standard and integrating Non-Fungible Token and Fungible Token technologies, proposes a spectrum securitization model to incentivize spectrum resource sharing and implements it on the Ethereum test net.

## I. Introduction

With the rapid development of 6G technology and the widespread adoption of 6G wireless networks, the demand for spectrum resources will reach an unprecedented level [1]. In the 6G era, competition with existing mobile communication new frequency band access rights holders caused by obtaining new frequency bands for mobile communication networks still exists [2]. Traditional static spectrum resource allocation schemes have low efficiency in spectrum utilization and can no longer meet the growing demand for spectrum resources. Dynamic spectrum resource sharing allows for flexible and real-time distribution of spectrum resources based on demand and usage patterns, optimizing the utilization of spectrum resources and enabling more efficient use of these resources [3]. However, to take advantage of dynamic spectrum resource allocation, a method to incentivize primary user (PU) to share spectrum resources is first needed.

Spectrum securitization, which converts spectrum access rights into tradable securities, offers increased revenue, risk mitigation, flexibility, and attracts investors, making spectrum sharing more financially attractive and sustainable [4]. Smart contract-based NFTs exhibit key traits like uniqueness, transparency, and immutability. These digital tokens are extensively utilized to represent ownership of both digital and physical assets [5] [6]. The use of smart contracts for managing spectrum resource transactions represented by NFTs can streamline the trading process, minimize intermediary fees, reduce the need for human intervention, and decrease the likelihood of disputes [7] [8] [9]. Using NFT to realize spectrum securitization has good prospects.

However, due to the indivisible nature of NFTs, they can only be traded as whole units. This characteristic, coupled with potentially higher prices, may lead to reduced liquidity for spectrum resources represented as NFTs. In November 2022, MarFernandez [10] proposed a F-NFT scheme. The F-NFT approach breaks down a complete NFT into smaller pieces. This allows multiple people to own parts of the same NFT at once. As a result, F-NFTs boost liquidity, spread out risk, offer flexible ownership options, and make trading easier. These benefits make NFTs more appealing to investors [11].

In spectrum securitization, multiple NFTs will be used to represent different spectrum assets. The F-NFT scheme is not suitable for simultaneously fragmenting multiple NFTs. In February 2024, the Pandora team [12] proposed an unofficial and experimental ERC404 standard. The ERC404 standard combines the ERC20 [13] standard and the ERC721 standard to provide the possibility for the fragmentation of multiple NFTs representing spectrum resources. The simplified diagram of the ERC404 standard for spectrum securitization is shown in Figure 1.1. The PU who is the owner of the smart contract mints FTs based on the number of spectrum assets. When a Secondary user (SU) possesses one Fungible Token (FT), the smart contract will mint an NFT. The NFT represents a spectrum asset by including spectrum asset's location and channel. In this way, the existing of the spectrum asset can be verified. When a FT owned by the SU is fragmented, the smart contract will burn one NFT in the SU's account. This is because the FT is owned by more than one SUs. In this way, the securitization of the spectrum is achieved by minting FTs. The spectrum resources are transparently displayed through NFTs.

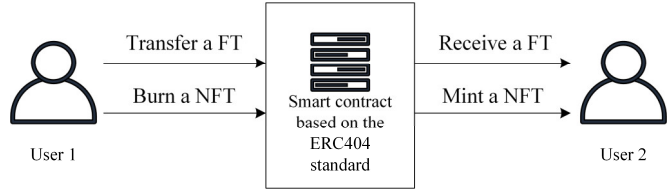

Figure 1.1 Simplified diagram of the ERC404 standard

To further encourage PUs to share spectrum resources, this scheme creates demand for FT by using FT for renting spectrum resources, instilling confidence in FT holders and supporting the sharing of spectrum resources. The ERC4907 standard [14] is an extension of the ERC721 standard, with its key feature being the separation of ownership and usage rights through role management for users and owners. This enhances the security of the leasing process. In this paper, we adopt the ERC404 standard as the NFT fragmentation solution to achieve spectrum resource securitization. Subsequently, the ERC4907 standard is used to design Non-Fungible Spectrum Token (NFST) for spectrum renting. SU can purchase FT to pay for renting NFST, allowing them to utilize the corresponding spectrum resources.

## II. SYSTEM MODEL

The model of the spectrum resource securitization system is shown in Figure 2.1. This system includes the following components: the spectrum management department, PUs, SUs, the spectrum resource tokenization module, the spectrum token leasing module.

a) The spectrum management department is responsible for the allocation, management, and regulation of spectrum resources.

b) PUs are the holders of spectrum resources.

c) SUs are those who have a demand for spectrum resources.

d) The spectrum resource tokenization module and the spectrum token leasing module are responsible for securitizing spectrum resources and minting tokens for spectrum resource sharing, respectively.

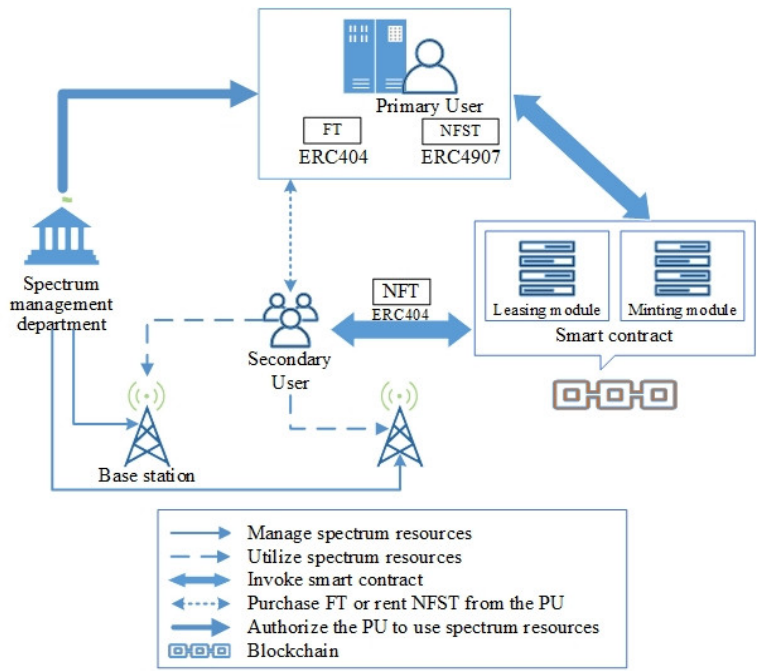

Figure 2.1 System model

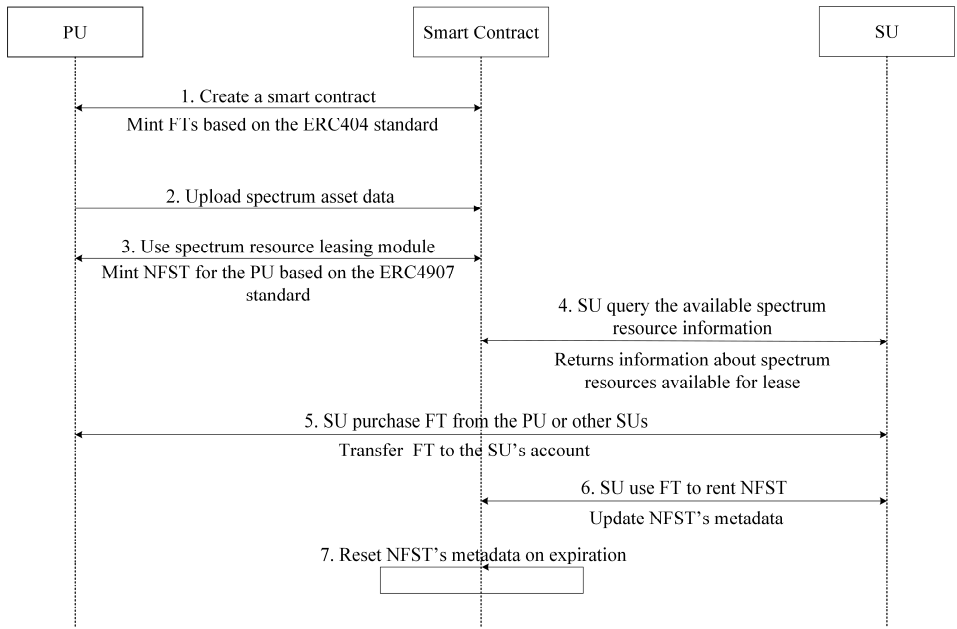

Figure 2.2 Spectrum tokenization and leasing modules

The processes for spectrum resource tokenization and leasing are shown in Figure 2.2. Provided the spectrum management department authorizes PUs to utilize the corresponding spectrum resources, the spectrum tokenization and leasing processes would proceed as follows:

1) The PU needs to use the spectrum resource tokenization module, for minting FT corresponding to the owned spectrum resources. FT can be purchased by anyone, anywhere, through decentralized smart contracts.

2) The PU uploads the corresponding spectrum resource information to the smart contract.

3) The PU uses the spectrum resource leasing module within the smart contract to mint NFST based on the ERC4907 standard. The NFST contains the owner's account address, spectrum resource information, the usage expiration time and the user's account address. The usage expiration time and the user's account address fields can be updated according to the request.

4) The SU calls the function in the smart contract to query available spectrum resources.

5) The SU purchases FT from PU or from other SU, and use FT to pay for renting NFST through the spectrum resource leasing module.

6) Upon successful rental, the NFST's user address will be updated to the SU's account address. The user's expire time is also updated. Other users can inspect the NFST's status to determine if the spectrum resource is currently occupied, thus preventing conflicts. The SU acquires the right to utilize the specified spectrum resources until the usage expiration time.

7) The system will reset the NFST's user address to zero when the usage period ends, making the spectrum resource available for new assignments.

## III. SMART CONTRACT DESIGN

We design smart contracts based on the ERC404 and ERC4907 standards to implement the spectrum resource securitization system module and the spectrum resource leasing module.

*A. Spectrum Resource Securitization*

After the spectrum management department authorizes PUs to utilize the corresponding spectrum resources, the PU calls the smart contract to mint FTs corresponding to the number of spectrum resources based on the ERC404 standard. In the spectrum resource trading system, the PU, as the owner of the smart contract, will not perform any minting or burning of NFTs when transferring FT. The FT minting function is shown in Figure 3.1, this function receives two parameters: the PU's address and the quantity of FT to be minted. This function is to transfer the newly minted FT to the PU's account and update the total supply of both FT and NFT. In this work, the unit of FT is wei ($1\ eth = 1 \times 10^{18}\ wei$).

```
Function mintFT:
    Input: recipient, amount_eth
    amount_wei ← amount_eth * (10 ** 18)
    balance[recipient] ← balance[recipient] + amount_wei
    totalSupply_FT ← totalSupply_FT + amount_wei
    totalSupply_NFT ← totalSupply_NFT + amount_eth
    emit TransferFT(address(0), recipient, amount_wei)
End Function
```

Figure 3.1 FT minting function

After the spectrum management department authorizes PUs to utilize the corresponding spectrum resources, the PU calls the smart contract to mint FTs corresponding to the number of spectrum resources based on the ERC404 standard. In the spectrum resource trading system, the PU, as the owner of the smart contract, will not perform any minting or burning of NFTs when transferring FT. The FT minting function is shown in Figure 3.1, this function receives two parameters: the PU's address and the quantity of FT to be minted. This function is to transfer the newly minted FT to the PU's account and update the total supply of both FT and NFT. In this work, the unit of FT is wei($1\ eth = 1 \times 10^{18}\ wei$).

B. *Upload Spectrum Resource Metadata*

```
Function updateChannelInfo:
    Input: channel, location
    if (isChannelUpload[channel] = false) then
        channelList.push(channel)
        channelInfo[channel].channel ← channel
        channelInfo[channel].location ← location
        return true
End Function
```

Figure 3.2 Spectrum metadata upload function

The PU needs to upload the spectrum resource metadata to the smart contract. This operation not only ensures that the spectrum resource securitization operation can proceed normally, but also provides transparency guarantee for the spectrum assets held by PU. The uploaded metadata contains the channel information and location information of the corresponding spectrum resources. The number of uploaded spectrum assets must be equal to the number of FTs minted and FT holders can view spectrum resources on the blockchain. The ERC404 standard ensures that the number of FTs does not exceed the number of spectrum resources. The pseudo code of the smart contract for uploading spectrum resource metadata is shown in Figure 3.2.

C. *Transfer of NFT*

```
Function transfer:
    Input: sender, recipient, amount_wei
    transferFT(sender, recipient, amount_wei)
    if (isPU[sender]=true && isPU[recipient]=false) then
        NFTNum_mint ← balance[recipient]_new - balance[recipient]_old
        mintNFT(recipient, NFTNum_mint)
        return true
    else if (isPU[sender]=false && isPU[recipient]=true) then
        NFTNum_burn ← balance[sender]_old - balance[sender]_new
        burnNFT(sender, NFTNum_burn)
        return true
    else if (isPU[sender]=false && isPU[recipient]=false) then
        NFTNum_burn ← balance[sender]_old - balance[sender]_new
        NFTNum_mint ← balance[recipient]_new - balance[recipient]_old
        burnNFT(sender, NFTNum_burn)
        mintNFT(recipient, NFTNum_mint)
        return true
End Function
```

Figure 3.3 FT transfer function

When the PU calls the ERC404-based transfer function, the smart contract first executes the FT transfer operation, and then determines whether to execute the NFT minting or destruction operation based on whether the sender or receiver is a PU and the FT balances changes of the sender and receiver. The smart contract pseudo code of the FT transfer function is shown in Figure 3.3.

Since this paper uses NFTs from the ERC404 standard to represent the spectrum resources. According to ERC404 standard, the PU, as the owner of the smart contract, will not perform any minting or burning of NFTs when transferring or receiving FTs. Once a SU acquires one or more FTs, the smart contract will be automatically triggered to select unoccupied spectrum resources in order from the uploaded spectrum resources to mint NFTs. For example, the order of uploading channel is: $C_1, C_2, \ldots, C_n$, assume $C_1$ is occupied and $C_2$ is not occupied, then $C_2$ will be used to mint NFT. However, since spectrum resources are a special type of resource, when users perform FT transfer operations under the ERC404 standard, it is necessary to discuss these operations in accordance with the spectrum management regulations of different regions.

a) Assuming that spectrum resources in a certain region are allowed to be traded, the NFT obtained by a SU when purchasing FTs will represent the ownership of the corresponding spectrum resource. If one of the SU's FTs is fragmented, the ownership of the corresponding spectrum resource will be lost as the associated NFT is burned.

b) Assuming that the spectrum resources in a certain place are not allowed to be traded, the NFT obtained by the SU when purchasing FTs can be used for transparent display of spectrum assets.

*D. Dynamic Spectrum Sharing*

```
Function rentNFSTByUser:
    Input: tokenIdOfNFST, renter
    if (isRented[tokenIdOfNFST] = true) then
        return false
    NFSTOwner ← nfstInfo[tokenIdOfNFST].Owner
    NFSTPrice ← nfstInfo[tokenIdOfNFST].price
    Duration ← nfstInfo[tokenIdOfNFST].duration
    transfer(renter, NFSTOwner, NFSTPrice)
    nfstInfo[tokenIdOfNFST].User ← renter
    nfstInfo[tokenIdOfNFST].expireTime ← Duration + block.timestamp
    emit RentNFSTByUser(tokenIdOfNFST, renter)
End Function
```

Figure 3.4 NFST leasing function

This paper proposes a scheme where FT is used to pay for renting NFST at a fixed price to facilitate spectrum resource sharing. The PU uses the spectrum resource leasing module within the smart contract based on the ERC4907 standard to mint NFST corresponding to the spectrum resources. The method for minting NFST can refer to literature [15]. Subsequently, the SU calls the NFST rental function and pays with FT to rent the NFST. The NFST rental function first retrieves the owner's address contained in the corresponding NFST, along with the usage duration and price set by the PU. It then transfers the FT corresponding to the NFST price from the buyer's address to the owner's address. Finally, it updates the user address of the NFST to the buyer's address and sets the expiration time for the NFST. The smart contract pseudo code of the NFST leasing function is shown in Figure 3.4.

## IV. EXPERIMENTAL RESULT

We deployed the spectrum securitization system in the Ethereum network. The Ethereum test net was built using Ganache, and smart contracts were deployed and called in the Remix platform. In the experiment, we choose one PU and two SUs. The information of the PU and SU is shown in Table I. The specific implementation process of the spectrum securitization system is as follows.

TABLE I. FT AND NFT INFO OF USERS

| User | Balance of FT (eth) | Balance of NFT |
|---|---|---|
| PU: 0x0aa7...1f60 | 8 | 0 |
| SU1: 0x50C0...9BCB | 0 | 0 |
| SU2: 0x408D…1F9F | 0 | 0 |
| SU3: 0x92E1…40c5 | 2 | 2 |
| SU4: 0x97d1…70D9 | 0 | 0 |

```
"from": "0xaf37e366e2ed4240fb039b9485f5ca1afda1849f",
"topic": "0x1c40d8e19403d18e6385fbc6a11a68d409eaca54f98bb7911ab02df7b2c61cb0",
"event": "TransferFT",
"args": {
        "0": "0x0000000000000000000000000000000000000000",
        "1": "0x0aa7652B45d957B9d2dE60AFbbD90b2DaD3d1f60",
        "2": "10000000000000000000",
        "_from": "0x0000000000000000000000000000000000000000",
        "_to": "0x0aa7652B45d957B9d2dE60AFbbD90b2DaD3d1f60",
        "_amount": "10000000000000000000"
}
```

Figure 4.1 FT minting result

The PU enters the number of spectrum assets held for minting FTs for the PU. The result of the FT minting is shown in Figure 4.1. In the figure, "from" refers to the smart contract's address, "topic" represents the hash value of the event signature, and "event" indicates the triggered "TransferFT" event. The "args" field contains three elements: 0, 1, and 2, corresponding to "_from", "_to", and "_amount", respectively. Here, "_from" represents the sender's address. Since the operation is minting FT, the sender's address corresponds to the zero address (0x000...0000). "_to" corresponds to the recipient's address. The transfer "_amount" is equal to 10 eths.

```
"from": "0xaf37e366e2ed4240fb039b9485f5ca1afda1849f",
"topic": "0x00f10c9f31b07f79511dfec1e765b78265a3bc77237673ebe07ac29ebb9d249c",
"event": "UploadChannelInfo",
"args": {
        "0": "Channel1",
        "1": "Location1",
        "_channel": "Channel1",
        "_location": "Location1"
}
```

Figure 4.2 Spectrum metadata upload result

The PU uploads the channel and location of the corresponding spectrum resource to the smart contract through the spectrum metadata upload function. The spectrum metadata upload result is shown in Figure 4.2.

```
"from": "0xaf37e366e2ed4240fb039b9485f5ca1afda1849f",
"topic": "0x3f29c9c4992d148d813fc5d021b82cae483c65cfa9be3a26cb1f8c80281a55e4",
"event": "TransferNFST",
"args": {
        "0": "0x0000000000000000000000000000000000000000",
        "1": "0x0aa7652B45d957B9d2dE60AFbbD90b2DaD3d1f60",
        "2": "1",
        "_from": "0x0000000000000000000000000000000000000000",
        "_to": "0x0aa7652B45d957B9d2dE60AFbbD90b2DaD3d1f60",
        "_tokenIdOfNFST": "1"
}
```

Figure 4.3 NFST minting result

The PU mint NFST by inputting the location and channel parameters. The result of the NFST minting is shown in Figure 4.3. Since this involves minting an NFST, "_from" is the zero address, "_to" is the PU's address, and "_tokenIdOfNFST" represents the tokenId of the newly minted NFST.

```
"from": "0xaf37e366e2ed4240fb039b9485f5ca1afda1849f",
"topic": "0x5cdd4d40c46fcff2ae2d1676edc7be8cb51f19ac0d6c006488370f5e2304d38b",
"event": "RentNFSTByOwner",
"args": {
        "0": "1",
        "1": "100000000000000000",
        "2": "86400",
        "_tokenIdOfNFST": "1",
        "_price": "100000000000000000",
        "_duration": "86400"
}
```

Figure 4.4 Results after PU rents out NFST

The PU call the "rent NFST" function in the smart contract, inputting the price and usage time parameters to rent out the NFST. The result of the PU renting out the NFST is shown in Figure 4.4. In the figure, "_tokenIdOfNFST" represents the tokenId of the NFST being rented by the PU, "_price" indicates the amount required to rent the NFST, which is 0.1 FT in this case, and "_duration" specifies the usage duration set by the PU for this NFST and its unit is seconds.

TABLE II. CASE 1

| User | FT amount before transfer (eth) | FT amount after transfer (eth) | Mint number of NFT | Burn number of NFT |
|---|---|---|---|---|
| PU: 0x0aa7...1f60 | 8 | 7 | 0 | 0 |
| SU1: 0x50C0...9BCB | 0 | 1 | 1 | 0 |

TABLE III. CASE 2

| User | FT amount before transfer (eth) | FT amount after transfer (eth) | Mint number of NFT | Burn number of NFT |
|---|---|---|---|---|
| PU: 0x0aa7...1f60 | 8 | 7.9 | 0 | 0 |
| SU2: 0x408D…1F9F | 0 | 0.1 | 0 | 0 |

TABLE IV. CASE 3

| User | FT amount before transfer (eth) | FT amount after transfer (eth) | Mint number of NFT | Burn number of NFT |
|---|---|---|---|---|
| SU3: 0x92E1…40c5 | 2 | 1 | 0 | 1 |
| SU4: 0x97d1…70D9 | 0 | 1 | 1 | 0 |

SU can purchase FT from the PU or other SUs. Three FT transferring cases are shown in Table II, Table III, Table IV. In the first case, the PU transfers one FT to SU1. After the transfer, the balance of the PU is 7 FTs. The balance of SU1 is 1 FTs, and one NFT is minted for SU1. In the second case, the PU transfers 0.1 FT to SU2. After the transfer, the balance of the PU is 7.9 FTs. The balance of SU2 is 0.1 FT which is less than 1 FT. There is no need to mint any NFTs for SU2. In the third case, SU3 transfers 1 FT to SU4. After the transfer, the balance of SU3 is 1 FT, one of SU3's NFT is burned. The balance of SU4 is 1 FT, one NFT is minted for SU4.

The information of the NFST is publicly visible across the network. SU can query the available NFST information in the spectrum securitization system to learn about the rental status of spectrum resources, helping to avoid conflicts between users.

```
"from": "0xaf37e366e2ed4240fb039b9485f5ca1afda1849f",
"topic": "0x1c40d8e19403d18e6385fbc6a11a68d409eaca54f98bb7911ab02df7b2c61cb0",
"event": "TransferFT",
"args": {
        "0": "0x408DD44B2c2Ebfdd0f9B66A448eEa7293B3c1F9F",
        "1": "0x0aa7652B45d957B9d2dE60AFbbD90b2DaD3d1f60",
        "2": "100000000000000000",
        "_from": "0x408DD44B2c2Ebfdd0f9B66A448eEa7293B3c1F9F",
        "_to": "0x0aa7652B45d957B9d2dE60AFbbD90b2DaD3d1f60",
        "_amount": "100000000000000000"
}

"from": "0xaf37e366e2ed4240fb039b9485f5ca1afda1849f",
"topic": "0xefbf87228b0beaf7ac94259c8cf96b15106728014028c2cd3047a22f127e9b28",
"event": "RentNFSTByUser",
"args": {
        "0": "1",
        "1": "0x408DD44B2c2Ebfdd0f9B66A448eEa7293B3c1F9F",
        "_tokenIdOfNFST": "1",
        "_renter": "0x408DD44B2c2Ebfdd0f9B66A448eEa7293B3c1F9F"
}
```

Figure 4.5 NFST rental result

As shown in Figure 4.5, the SU2 first transfers 0.1 FT to the PU's account to rent the NFST. Subsequently, the smart contract will then set the user address in NFST to the address of SU2 and SU2 can use the corresponding spectrum resources.

V. CONCLUSION

This article designs a smart contract based on the ERC404 standard and ERC4907 standard to securitize spectrum resources and encourage PUs to share spectrum resources. The ERC404 standard ensures that the number of FTs does not exceed the number of spectrum resources and that spectrum resources remain transparent and visible. SU uses FT to lease NFST in accordance with the ERC 4907 standard, thus obtaining the temporary right to use the spectrum. The rental of NFST enhances the liquidity of FT, generating revenue for the system and encouraging PUs to be more willing to share spectrum resources, thereby establishing a spectrum resource sharing ecosystem between PUs and SUs.

ACKNOWLEDGMENT

Guangdong Province Graduate Education Innovation Pro-gram Project [2024JGXM_163], Shenzhen University Graduate Education Reform Research Project [SZUGS2023JG02], Shenzhen University Teaching Reform Research Project [JG2023097],

## REFERENCES

[1] Cuellar D, Sallal M, Williams C. BSM-6G: Blockchain-Based Dynamic Spectrum Management for 6G Networks: Addressing Interoperability and Scalability[J]. IEEE Access, 2024.

[2] Matinmikko-Blue M, Yrjölä S, Ahokangas P. Spectrum management in the 6G era: The role of regulation and spectrum sharing[C]//2020 2nd 6G Wireless Summit (6G SUMMIT). IEEE, 2020: 1-5.

[3] Jayaweera S K, Vazquez-Vilar G, Mosquera C. Dynamic spectrum leasing: A new paradigm for spectrum sharing in cognitive radio networks[J]. IEEE transactions on vehicular technology, 2010, 59(5): 2328-2339.

[4] Balzacq T. Securitization theory[J]. How security problems emerge and dissolve, 2011.

[5] William Entriken (@fulldecent), Dieter Shirley <dete@axiomzen. co>, Jacob Evans <jacob@dekz.net>, Nastassia Sachs <nastassia. sachs@protonmail.com>, "ERC-721: Non-Fungible Token Standard," Ethereum Improvement Proposals, no. 721, January 2018. [Online serial]. Available: https://eips.ethereum.org/EIPS/eip-721.

[6] Wang Q, Li R, Wang Q, et al. Non-fungible token (NFT): Overview, evaluation, opportunities and challenges[J]. arXiv preprint arXiv:2105.07447, 2021.

[7] Shao X, Cao P, Wang S, et al. Non-Fungible Token Enabled Spectrum Sharing for 6G Wireless Networks[C]//2023 IEEE Globecom Workshops (GC Wkshps). IEEE, 2023: 1075-1080.

[8] Perera L, Ranaweera P, Wang S, et al. Spect-NFT: Non-Fungible Tokens for Dynamic Spectrum Management[J].

[9] Zhou Z, Chen X, Zhang Y, et al. Blockchain-empowered secure spectrum sharing for 5G heterogeneous networks[J]. IEEE Network, 2020, 34(1): 24-31.

[10] MarFernandez.eth. "The entire f-NFT space explained." Mirror, 5 Jun. 2024, https://mirror.xyz/marfernandez.eth/s4fgUVp34_1q3zTT64qfPjCm97cLpFTy5B60vbQ78Kg.

[11] Choi W, Woo J, Hong J W K. Fractional non‐fungible tokens: Overview, evaluation, marketplaces, and challenges[J]. International Journal of Network Management, 2024: e2260.

[12] Pandora Labs®. https://www.pandora.build/. Accessed 5 Jun. 2024.

[13] Fabian Vogelsteller <fabian@ethereum.org>, Vitalik Buterin <vitalik.buterin@ethereum.org>, "ERC-20: Token Standard, " Ethereum Improvement Proposals, no. 20, November 2015. [Online serial]. Available: https://eips.ethereum.org/EIPS/eip-20.

[14] Anders (@0xanders), Lance (@LanceSnow), Shrug <shrug@ emojidao.org>, "ERC-4907: Rental NFT, an Extension of EIP-721," Ethereum Improvement Proposals, no. 4907, March 2022. [Online serial]. Available: https://eips.ethereum.org/EIPS/eip-4907.

[15] Ye L, Chen B, Shivanshu S, et al. Dynamic Spectrum Sharing Based on the Rentable NFT Standard ERC4907[C]//2024 13th International Conference on Communications, Circuits and Systems (ICCCAS). IEEE, 2024: 503-507.